\documentclass[epj,referee]{svjour}
\usepackage{amssymb}
\usepackage{latexsym}
\usepackage[dvipdfmx]{graphicx}
\begin{document}
\title{
Criticality of the magnon-bound-state hierarchy for 
the quantum Ising chain
with the long-range interactions
}
\author{Yoshihiro Nishiyama  
}                     
\offprints{}          
\institute{Department of Physics, Faculty of Science,
Okayama University, Okayama 700-8530, Japan}
\date{Received: date / Revised version: date}
%
\abstract{
The quantum Ising chain with the interaction
decaying as a power law $1/r^{1+\sigma}$ of the distance between spins
$r$
was investigated numerically.
A particular attention was paid to the
low-energy spectrum,
namely,
the single-magnon and two-magnon-bound-state
masses, $m_{1,2}$, respectively,
in the ordered phase.
It is anticipated that 
for each $\sigma$, the scaled bound-state mass 
$m_2/m_1$ should take a universal constant 
(critical amplitude ratio)
in the vicinity of the critical point.
In this paper,
we calculated
the amplitude ratio $m_2/m_1$
with the
exact diagonalization method,
which yields the spectral information such as $m_{1,2}$ directly.
As a result,
we found that the scaled mass $m_2/m_1$
exhibits a non-monotonic dependence on $\sigma$;
that is, the bound state is stabilized by an intermediate
value of $\sigma$.
Such a feature is accordant with a recent observation 
based on the
non-perturbative-renormalization-group method.
} 
\maketitle

\section{\label{section1}Introduction}

\begin{figure}
\resizebox{0.5\textwidth}{!}{%
\includegraphics{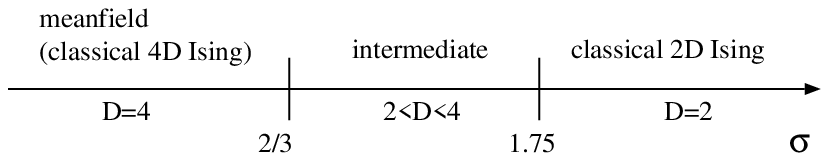}  }%
\caption{
\label{figure1}
The quantum Ising chain
with the algebraically decaying interactions 
$\propto1/ |i-j|^{1+\sigma}$,
Eq.
(\ref{Hamiltonian}),
exhibits the order-disorder phase transition.
The 
 criticality (universality class)
depends on the fall-off exponent $\sigma$,
and the singularity is classified into three regimes
\cite{Defenu17}.
For small $\sigma<2/3$, the phase transition belongs to the mean-field type,
namely, the $D=4$-classical-Ising universality class.
For large $\sigma>1.75$, on the contrary,
it is identical to that of the $D=2$ Ising model.
In the intermediate regime $2/3 < \sigma < 1.75$, the singularity is 
controlled by $\sigma$ continuously,
and correspondingly,
the fractional dimensionality changes within
$2<D<4$ \cite{Angelini14}.
At the boundaries, paticularly, at $\sigma=1.75$,
there appear notorious logarithmic corrections
\cite{Luijten02,Brezin14,Defenu15,Fey16}.
}
\end{figure}

The O$(N)$-symmetric classical spin model with the long-range interactions 
has been investigated
both theoretically 
\cite{Fisher72,Sak73,Luijten02,
Picco07,Blanchard13,Grassberger13,Gori17,Angelini14,
Joyce66,Brezin14,Defenu15,Defenu16,Goll18,%
FloresSola17,Horita17,Sun96,%
Humeniuk16}
and experimentally
\cite{Ellman91,Britton12,Islam13,Bohnet16,Zhang17,%
Richerme14,Jurcevic14,Paz13,Browaeys16,Moses17,Schauss12}.
A notable feature is that the power of the algebraic decay
affects the 
criticality of the order-disorder phase transition \cite{Fisher72,Sak73}.
Meanwhile,
an extention to the quantum-mechanical version was made
\cite{Laflorencie05,Dutta01,Defenu17,Fey16,Sandvik10,Koffel12}.
As for the quantum Ising chain with the long-range
interactions
decaying as a power law $1/r^{1+\sigma}$ of the distance between spins 
$r$, there should appear three distinctive 
types of criticalities \cite{Defenu17};
see Fig. \ref{figure1}.
For $\sigma < 2/3$, the 
criticality belongs to the
mean-field type;
namely, the singularity is
identical to that of the four dimensional ($D=4$)
short-range classical Ising model.
On the contrary, in $\sigma > 1.75$, the long-range interaction becomes
irrelevant, 
and eventually, the criticality reduces to the
$D=2$-short-range-classical-Ising universality class.
In the intermediate regime 
$2/3 < \sigma < 1.75 $,
the singularity depends soothly on $\sigma$.
Accordingly,
the fractional dimensionality ranges  
within $2<D<4$ \cite{Gori17,Angelini14,Joyce66,Defenu17}.
At both boundaries, particularly,
at $\sigma=1.75$,
there emerge notorious logarithmic corrections
\cite{Luijten02,Brezin14,Defenu15,Fey16},
and the details as to the end-point singularities 
are controversial
\cite{Picco07,Blanchard13,Grassberger13}.
It has to be mentioned that the 
above-mentioned features 
are not a mere theoretical interest,
because such an adjustable
algebraic-decay rate $\sigma$
is realized experimentally
\cite{Britton12,Islam13,Richerme14}.

\begin{figure}
\resizebox{0.5\textwidth}{!}{%
\includegraphics{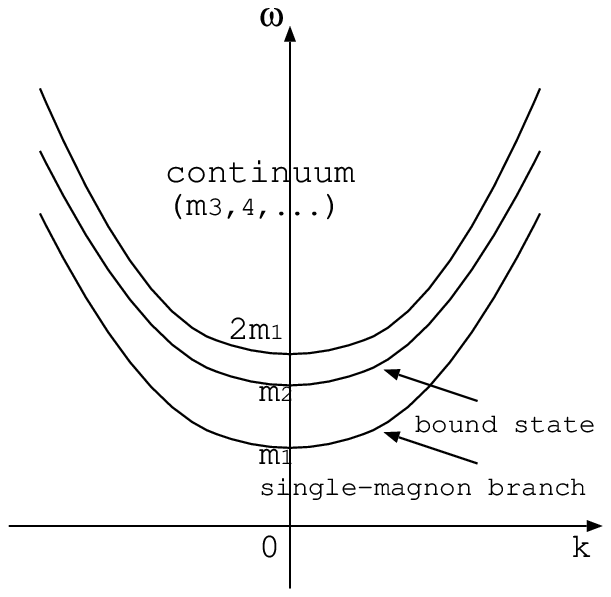}  }%
\caption{
\label{figure2}
A schematic drawing for the low-lying spectrum
of the quantum Ising model
in the ordered phase is presented.
At the zone center $k=0$,
there open 
the single-magnon- and two-magnon-bound-state-mass gaps, $m_{1,2}$,
respectively.
Above $2 m_2$, there extends
a continuum, and the magnon-bound-state hierarchy
$m_{3,4,\dots}$ should be embedded within the continuum.
The bound state is stabilized around 
$D\approx 3$,
as shown in Fig. \ref{figure3}.
}
\end{figure}

\begin{figure}
\resizebox{0.5\textwidth}{!}{%
\includegraphics{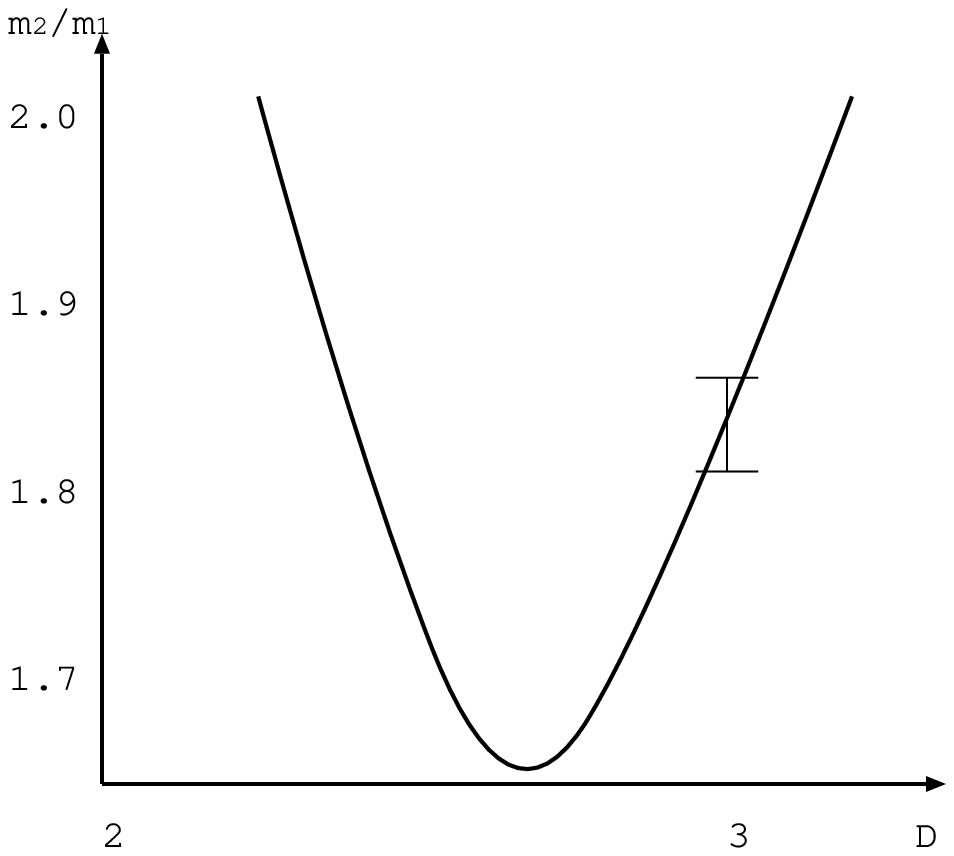}  }%
\caption{
\label{figure3}
According to the non-perturbative renormalization-group
analysis \cite{Rose16},
the scaled bound-state mass 
$m_2/m_1$
exhibits a non-monotonic dependence on the 
fractional dimensionality $D$ of the short-range 
{\em classical} Ising model,
namely, the $(D-1)$-dimensional {\em quantum} Ising model with the short-range interactions.
As a reference, 
a plot
$m_2/m_1=1.82(2)$ at $D=3$ \cite{Rose16},
which is of particular importance,
is indicated.
}
\end{figure}

The criticality chart, Fig. \ref{figure1},
differs from
 that of the classical counterpart
\cite{Fisher72,Sak73}.
The difference comes from the fact that
the real-space and imaginary-time directions 
are anisotropic for the quantum long-range criticality;
the anisotropy 
is characterized by the dynamical 
critical exponent $z$
\cite{Defenu17},
and accordingly, the set of scaling relations
should be remedied.
Aiming to elucidate the quantum-mechanical character of this problem,  
we shed light on the low-lying spectrum.
In Fig. \ref{figure2},
we present a
schematic drawing for the dispersion relations
as to the low-lying elementary excitations
in the ordered phase \cite{Rose16}.
Here, the symbol $m_{1(2)}$
denotes the single-magnon (two-magnon-bound-state)
mass.
Above $2 m_1$, there extends
a continuum, and the series of bound states $m_{3,4,\dots}$
should be embedded within the continuum \cite{Coldea10,Caselle99}.
The bound-state hierarchy,
namely, the scaled mass
$m_{2,3,\dots}/m_1$,
displays a universal character in the vicinity of the critical point.
As a matter of fact,
according to the non-perturbative
renormalization group \cite{Rose16},
the scaled mass $m_2/m_1$ 
exhibits a non-monotonic dependence on 
$D$
(fractional dimensionality); 
see Fig. \ref{figure3}.
That is, an intermediate value of $D$
stabilizes the bound state.
Rather intriguingly,
the hierarchy $m_{2,3,\dots}/m_1$ is relevant to the
high-energy physics, that is, the
gluon-bound-state spectrum 
(the so-called glueball spectrum) for the
gauge field theory \cite{Agostini97,Fiore03}.
The universal values $m_{2,3}/m_1$ for the Ising model 
were explored extensively
 in this context \cite{Caselle99}.

In this paper, 
we investigate
the scaled
two-magnon-bound-state
mass gap $m_2/m_1$
for the spin-$S=1$ Ising chain with the long-range interactions
(\ref{Hamiltonian}).
The extended spin $S=1$ 
permits
 us to deal with
the generalized (quadratic) interactions $(D_s,\gamma_2)$,
which admit
a clear signature for the bound-state mass
(such as the plateau in Fig. \ref{figure6} mentioned afterward)
in the finite-size data.
We employed the exact diagonalization method,
which enables us to calculate the spectral properties such as $m_{1,2}$
directly without resorting to the inverse Laplace transformation;
see Appendix B of Ref. \cite{Gazit13}.
In fairness, it has to be mentioned
that recently,
the dynamical properties for
 the long-range systems were studied
with
the exact diagonalization and the variational-matrix-product methods
\cite{Homrighausen17,Frerot18,Vanderstraeten18}.

To be specific,
the
Hamiltonian for the quantum spin-$S=1$ Ising chain
with the long-range interactions
is given by
\begin{eqnarray}
\label{Hamiltonian}
{\cal H} & =  &
 - \frac{1}{\cal N}\sum_{i \ne j} J_{ij} S^z_i S^z_j
   + D_s \sum_i  (S^z_i)^2    \nonumber \\
      & &
  - \Gamma 
  \sum_i (\frac{S^+_i}{\sqrt{2}} + \frac{\gamma_2}{2}(S^+_i)^2 +h.c.)
,
\end{eqnarray}
with the quantum spin-$S=1$ operator 
$S^{\pm,z}_i$ placed at each one-dimensional lattice 
point,
$i=1,2,\dots,L$.
Here,
the periodic boundary condition is imposed.
The summation
$\sum_{i\ne j}$ runs over all possible $i$-$j$ pairs,
and the symbol
$J_{ij}$ denotes the algebraically decaying
interaction,
$J_{ij}=1/\sin(\pi|i-j|/L)^{1+\sigma}$,
parameterized by $\sigma$.
The Kac-normalization factor 
${\cal N}$
\cite{Homrighausen17,Vanderstraeten18}
is given by 
${\cal N}=L^{-1}\sum_{i\ne j}\sin(\pi|i-j|/L)^{-1-\sigma}$.
The summation $\sum_i$ runs over all spins $i=1,2,\dots,L$,
and the single-ion anisotropy $D_s$ is incorporated.
The transverse magnetic field
$\Gamma$, together with its quadratic variant $\gamma_2$,
drives the ferromagnetic state to the disordered phase.
These
redundant interaction parameters $(D_s,\gamma_2)$ are 
 tuned so as to attain a clear indication for the 
magnon bound state.

The rest of this paper is organized as follows.
In the next section,
we present the simulation results for
 the long-range Ising chain (\ref{Hamiltonian}).
The simulation algorithm is explained as well.
In the last section,
we address the summary and discussions.

\section{\label{section2}Numerical results}

In this section,
we present the numerical results
for 
the long-range quantum Ising chain (\ref{Hamiltonian}).
We employed the
exact diagonalization method,
which enables us to calculate the mass gaps $m_{1,2}$ directly.
The numerical diagonalization was performed
within the restricted Hilbert space
specified by the quantum numbers
such as 
the zero-momentum ($k=0$) and
the spin-inversion ($S^z_i \to -S^z_i $)
parity index, 
$\pm$.
In a preliminary survey, we found that
the single-magnon mass $m_1$ 
belongs to the sector
\begin{equation}
m_1=E_{1}^+ -E_{0}^+ ,
\end{equation}
with the ground-state (first-excited) energy 
$E_{0}^+$
($E_1^+$) with the parity index $+$.   
Similarly, the two-magnon-bound-state mass $m_2$ is identified as
either
\begin{equation}
\label{specification1}
m_2=E_3^+-E_0^+ ,
\end{equation}
or
\begin{equation}
\label{specification2}
m_2=E_2^--E_0^+ ,
\end{equation}
for small- and large-$\sigma$ regimes, respectively;
that is, the character of $m_2$ changes for respective
regimes,
as suggested by Fig. \ref{figure3} \cite{Rose16}.

\subsection{\label{section2_1}Preliminary survey:
Scaling behavior for the single-magnon mass $m_1$}

As a preliminary survey, in this section,
we investigate
the scaling behavior for the 
single-magnon mass $m_1$.
To this end,
we make use of
the scaling theory developed in
Ref. \cite{Defenu17}.
The interaction parameters are set to
$(D_s,\gamma_2)=(0,2)$
throughout this section.

\begin{figure}
\resizebox{0.5\textwidth}{!}{%
\includegraphics{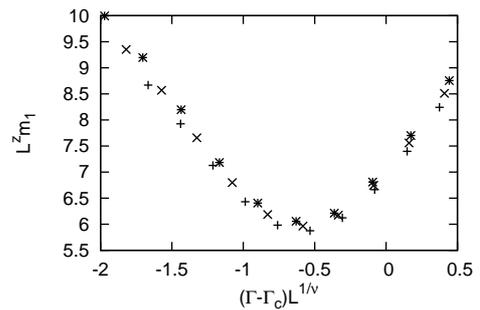}  }%
\caption{
\label{figure4}
The scaling plot, 
$(\Gamma-\Gamma_c)L^{1/\nu}$-$L^z m_1$,
with 
the
critical point $\Gamma_c=0.527$,
the reciprocal correlation-length critical exponent $1/\nu=0.84$,
and the dynamical critical exponent
$z=0.60$
is presented for $\sigma=1.2$
and various
system sizes, 
($+$) $L=18$,
($\times$) $20$,
and
($*$) $22$;
see text for details.
Here, the interaction parameters are set to $(D_s,\gamma_2)=(0,2)$.
}
\end{figure}

In Fig. \ref{figure4}, we present the scaling plot,
$(\Gamma-\Gamma_c)L^{1/\nu}$-$L^z m_1$, with
the critical point $\Gamma_c=0.527$, 
the reciprocal correlation-length critical exponent 
$1/\nu=0.84$, and
the dynamical critical exponent $z=0.60$
for the fixed $\sigma=1.2$ and various system sizes,
($+$) $L=18$,
($\times$) $20$,
and ($*$) $22$.
Here,
the critical point $\Gamma_c=0.527$
was extrapolated 
via the
least-squares fit 
for the $L^{-1}$-$\Gamma_c(L)$ data with
$L=18,20,22$,
and
the approximative critical point 
$\Gamma_c(L)$ was determined by 
the condition
$\partial_\Gamma m_1|_{\Gamma=\Gamma_c(L)} = 0$
for each $L$.
Similarly, the dynamical critical exponent $z=0.60$ was determined
via
 the least-squares fit for 
$(\frac{L+(L+2)}{2})^{-1}$-$z(L,L+2)$  with $L=16,18,20$,
and
the approximative dynamical critical exponent
$z(L,L')$
is given by
\begin{equation}
z(L,L')=-\frac{
\ln m_1(L)|_{\Gamma=\Gamma_c(L)}-\ln m_1(L')|_{\Gamma=\Gamma_c(L')}}
{\ln L-\ln L'}
 .
\end{equation}
As for the reciprocal correlation-length critical exponent 
$1/\nu=0.84$,
we made use of the existing value as addressed in Figs. 2 and 3
of Ref. 
\cite{Defenu17}.

From Fig. \ref{figure4},
we notice that
the data collapse into a scaling curve
satisfactorily.
The 
single-magnon gap $m_1$ opens in the ordered phase $\Gamma-\Gamma_c<0$.
The gap $m_1$ sets a fundamental energy scale in the subsequent analyses
in
Sec. \ref{section2_2} and \ref{section2_3}.

\begin{figure}
\resizebox{0.5\textwidth}{!}{%
\includegraphics{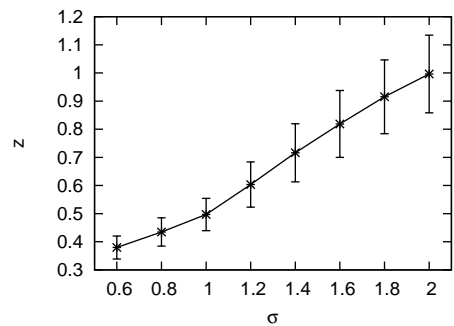}  }%
\caption{
\label{figure5}
The dynamical critical exponent $z$
is presented for various values of 
the algebraically-fall-off exponent $\sigma$;
see text for details.
Here, the interaction parameters are set to $(D_s,\gamma_2)=(0,2)$.
}
\end{figure}

\begin{figure}
\resizebox{0.5\textwidth}{!}{%
\includegraphics{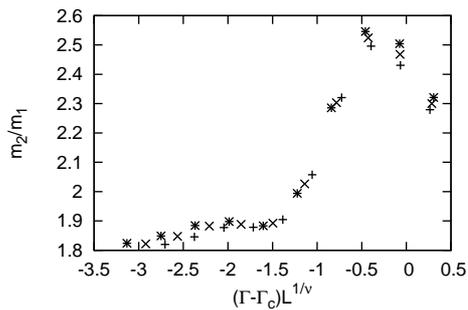}  }%
\caption{
\label{figure6}
The scaling plot,
$(\Gamma-\Gamma_c)L^{1/\nu}$-$m_2/m_1$,
with the critical point $\Gamma_c=0.568$
and
the reciprocal correlation-length critical exponent 
$1/\nu=0.73$
is presented for 
 $\sigma=0.8$
and various
system sizes, 
$(+)$ $L=18$,
$(\times)$ $20$,
and 
$(*)$ $22$.
Here, the interaction parameters are set to $(D_s,\gamma_2)=(-0.1,2)$.
The height of the plateau  $m_2/m_1 \approx 1.9$ 
indicates an amplitude ratio;
the extrapolated one is plotted in Fig. \ref{figure7}.
As the system size enlarges,
the shoulder around 
$(\Gamma-\Gamma_c)L^{1/\nu} \approx -2.5$
extends leftward,
suggesting the stabilization of the bound state
in the thermodynamic limit.
}
\end{figure}

Carrying out simular analyses for various values of $\sigma$,
we obtained the dynamical critical exponent
$z$. The result is presented
in Fig. \ref{figure5}.
Here,  as an indicator for the error,
we accept
the deviation between the different 
least-squares-fit analyses
(abscissa scales), namely, 
the
$(\frac{L+(L+2)}{2})^{-1}$-
and
$(\frac{L+(L+2)}{2})^{-2}$-based extrapolation schemes.
In the large-$\sigma$ side ($D \to 2$),
the symmetry between the real-space and imaginary-time directions 
restores, and the
dynamical critical exponent reflects the recovery,
$z \to 1$.
On the contrary, in the small-$\sigma$ side,
these spaces become asymmetric,
 $z < 1$.
We stress that our simulation 
covers these two extreme cases
with the interaction parameters fixed to 
$(D_s,\gamma_2)=(0,2)$.

A few remarks are in order.
First,
the data at both boundaries,
$\sigma=2/3$ and $1.75$, should 
suffer from corrections to scaling
\cite{Luijten02,Brezin14,Defenu15,Fey16};
particularly,
the latter one
at  $\sigma=1.75$ has arisen controversies 
\cite{Picco07,Blanchard13,Grassberger13}
as to the details of the end-point singularity.
Because our main concern is the midst regime $\sigma \approx 1$,
we do not pursue the details any further.
Second,
our data appear to obey
an approximative formula
\cite{Defenu17}
\begin{equation}
\label{z_formula}
z=\sigma/2 .
\end{equation}
Actually,
as mentioned above, we obtained $z=0.60$ for $\sigma=1.2$;
the result seems to 
agree with
 the formula, Eq. (\ref{z_formula}).
The validity of Eq. (\ref{z_formula}) is not guaranteed for large $\sigma$
\cite{Defenu17} nonetheless.
Last,
the simulation was performed with the interaction
parameters
fixed to
$(D_s,\gamma_2)=(0,2)$.
In the subsequent sections,
we adjust the interaction parameters
for the small- and large-$\sigma$ regimes separately
in order to attain a clear indication for $m_2/m_1$.

\subsection{\label{section2_2}Scaled two-magnon-bound-state mass $m_2/m_1$: Small-$\sigma$ side}

In this section,
we analyze the scaled two-magnon-bound-state mass $m_2/m_1$
in the small-$\sigma$ side.
Here, 
the parameters are fixed to 
$(D_s,\gamma_2)=(-0.1,2)$ so as to
attain an appreciable plateau for $m_2/m_1$.

\begin{figure}
\resizebox{0.5\textwidth}{!}{%
\includegraphics{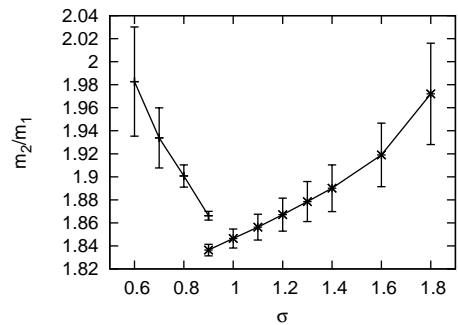}  }%
\caption{
\label{figure7}
The scaled bound-state mass $m_2/m_1$ is presented
for various values of the algebraically-fall-off exponent $\sigma$.
The plots, ($+$) and ($\times$),
are determined in Sec. \ref{section2_2} and \ref{section2_3},
respectively.
}
\end{figure}

In Fig. \ref{figure6},
we present the scaling plot, 
$(\Gamma-\Gamma_c)L^{1/\nu}$-$m_2/m_1$,
with the critical point 
$\Gamma_c=0.568$ and the reciprocal 
correlation-length
critical exponent $1/\nu=0.73$
for $\sigma=0.8$ and various system sizes, 
($+$) 
$L=18$,
($\times$) $20$,
and 
($\times$) $22$;
here, the critical point 
$\Gamma_c=0.568$ was determined with the same scheme as
in Sec. \ref{section2_1},
and 
the reciprocal critical exponent 
$1/\nu=0.73$ is taken from Figs. 
2 and 3
of Ref. \cite{Defenu17}.
We see a plateau with the height
$m_2/m_1 \approx 1.9$ in the ordered phase 
$(\Gamma-\Gamma_c)L^{1/\nu} \approx -2 (< 0)$.
We arrived at $m_2/m_1=1.895(10)$ via
the least-squares fit for
$L^{-1}$-$m_2/m_1|_{\Gamma=\bar{\Gamma}(L)}$ with $L=18,20,22$;
here, 
the location of the plateau
(shallow valley) floor $\bar{\Gamma}(L)$
satisfies
$\partial_\Gamma (m_2/m_1)|_{\Gamma=\bar{\Gamma}(L)} =0$
for each $L$,
and as an indicator for the error,
we accept the deviation
between the different least-squares-fit analyses (abscissa scales),
namely, the
$L^{-1}$- and $L^{-2}$-based extrapolation schemes.
Carrying out simular analyses for various values of $\sigma$,
we obtained the estimates
as
indicated by
the symbol ($+$) in Fig. \ref{figure7}.

We address a number of remarks.
First, for exceedingly large $\sigma$,  
the plateau width shrinks, and eventually, 
the plateau disappears.
Such a feature 
suggests that the bound state 
(belonging to the ($+$) branch)
is no longer supported
by the long-range interactions with  exceedingly large $\sigma$.
Second, the shoulder around 
$(\Gamma-\Gamma_c)L^{1/\nu}\approx -2.5$ grows, as the system size $L$ enlarges.
Such a feature indicates the
stability of the bound state.
Last,
around the boundary $\sigma=2/3$,
the simulation data suffer from the systematic errors,
as noted in Ref. \cite{Fey16}.
Because our concern is to survey the bound-state stabilization around $\sigma \approx 1$,
we do not pursue the details any further.

\subsection{\label{section2_3}Scaled two-magnon-bound-state mass $m_2/m_1$: Large-$\sigma$ side}

In this section,
we analyze the scaled two-magnon-bound-state mass $m_2/m_1$
in the large-$\sigma$ side.
Here, the parameters are fixed to 
$(D_s,\gamma_2)=(0.2,0.25)$.

\begin{figure}
\resizebox{0.5\textwidth}{!}{%
\includegraphics{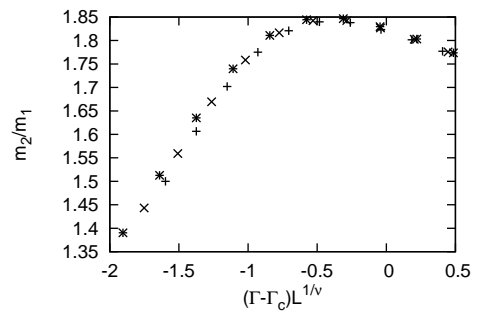}  }%
\caption{
\label{figure8}
The scaling plot,
$(\Gamma-\Gamma_c)L^{1/\nu}$-$m_2/m_1$,
with the critical point $\Gamma_c=0.913$
and
the reciprocal correlation-length critical exponent
$1/\nu=0.88$
is presented for 
 $\sigma=1.3$
and various
system sizes,
$(+)$ $L=18$,
$(\times)$ $20$,
and 
$(*)$ $22$.
Here, the interaction parameters are set to $(D_s,\gamma_2)=(0.2,0.25)$.
The hilltop height $m_2/m_1 \approx 1.85$ 
indicates an amplitude ratio;
the extrapolated one is plotted in Fig. \ref{figure7}.
}
\end{figure}

In Fig. \ref{figure8},
we present the scaling plot, $(\Gamma-\Gamma_c)L^{1/\nu}$-$m_2/m_1$,
with the critical point $\Gamma_c=0.913$ and the reciprocal 
correlation-length critical exponent $1/\nu=0.88$
for $\sigma=1.3$ and various system sizes,
($+$) $L=18$,
($\times$) $20$, and  ($*$) $22$;
here, the critical point $\Gamma_c=0.913$ was
determined with the same scheme as in Sec. \ref{section2_1}, and the exponent 
$1/\nu=0.88$ is taken from Figs. 2 and 3 of 
Ref. \cite{Defenu17}.
The hilltop 
height $m_2/m_1\approx 0.85$ indicates the scaled bound-state mass.
Via the
least-squares fit for
 $L^{-1}$-$m_2/m_1|_{\Gamma=\tilde{\Gamma}(L)}$ 
with $L=18,20,22$,
we obtained
an estimate
$m_2/m_1=1.879(17)$;
here, the hilltop location $\tilde{\Gamma}(L)$
satisfies
$\partial_\Gamma(m_2/m_1)|_{\Gamma=\tilde{\Gamma}(L)}=0$
for each $L$,
and as an indicator for the error,
we accept the deviation
between the different least-squares-fit analyses
(abscissa scales), namely,
the
$L^{-1}$- and $L^{-2}$-based extrapolation schemes.
Carrying out simular analyses for
a variety of $\sigma$,
we obtained a series of results
as
indicated 
by the symbol
($\times$) in Fig. \ref{figure7}.

We address a number of remarks.
First, for exceedingly small $\sigma$,
the hilltop location $\tilde{\Gamma}$
shifts into the disordered phase,
and the branch terminates.
Last, the data around the boundary $\sigma=1.75$ should 
suffer from
the systematic errors 
\cite{Luijten02,Brezin14,Defenu15}.

\subsection{\label{section2_4}
Comparison with the preceeding results via
the
$\sigma \leftrightarrow D$ relation
\cite{Angelini14}
}

In this section,
we make a comparison 
with the preceding results
such as the fixed-$D=3$ analyses
\cite{Caselle99,Dusuel10,Nishiyama14}
and the non-perturbative renormalization group
for $^\forall D$
\cite{Rose16}.
In order to establish a relationship between
them and ours, 
we rely on the 
$\sigma \leftrightarrow D$ relation
\cite{Angelini14,Defenu17}
\begin{equation}
\label{sDrelation}
D = 2/\sigma +1 ,
\end{equation}
which is validated for small-$\sigma$ (large-$D$) regime;
more sophisticated formulas \cite{Defenu17}
do not take such a closed expression.

The overall features of Fig. \ref{figure7} 
and \ref{figure3} \cite{Rose16}
resemble 
each other;
actually,
the magnon bound state is stabilized by an intermediate value
of $\sigma$ and $D$, respectively. 
Such stabilization of the bound-state mass
is
 captured by neither  mean-field theory ($D=4$) nor
 free-fermion picture  ($D=2$).

A number of remarks are in order.
First,
we consider the case $D=3$.
This case corresponds to $\sigma=1$ according to the 
$\sigma \leftrightarrow  D$ relation
(\ref{sDrelation}).
Our result indicates $m_2/m_1=1.845(10)$ at $\sigma=1$
(along side of the ($\times$) branch).
On the one hand,
by means of the Monte Carlo \cite{Caselle99}, 
series expansion \cite{Dusuel10}, 
and exact diagonalization \cite{Nishiyama14} methods,
the estimates, $m_2/m_1=1.83(3)$,
$1.81$,
and $1.84(1)$, respectively, were obtained for the fixed-$D=3$ systems.
Additionally,
the $^\forall D$ non-perturbative-renormalization-group method
yields $m_2/m_1=1.82(2)$ at $D=3$ \cite{Rose16}.
Our result appears to be consistent with these preceeding elaborated
analyses.
According to the super-universality idea,
the long-  and the short-range models are related with 
a relation such as Eq. (\ref{sDrelation}) at least for the large-$D$
side
\cite{Angelini14}.
Our data suggest that down to $D=3$, the relationship
 is retained even for the spectral properties such as $m_2/m_1$.
Second,
we turn to considering the case, $D \ne 3$.
The non-perturbative renormalization group 
\cite{Rose16} predicts that the
minimum point locates around $D=2.7$-$2.8$,
where the scaled mass takes  $m_2/m_1=1.65$-$1.7$.
On the contrary, our result resolves neither appreciable deviation of the minimum point
from $D=3$ ($\sigma=1$) nor notable drop of $m_2 / m_1$ in the small-$D$ side.
The discrepancy may be attributed to the breakdown of 
a naive $\sigma \leftrightarrow D$ correspondence 
for such
 small-$D$ regime.
Last, 
we provide a brief overview on  the $\sigma \leftrightarrow D$ relation
(\ref{sDrelation}).
In the course of the studies \cite{Gori17,Angelini14,Defenu15}, the 
concept of the
$\sigma \leftrightarrow D$ relation 
has been developed.
As for the quantum-mechanical system,
there was reported
a refined formula  
$D=(2-\eta_{SR}(D))\frac{1+z(\sigma)}{\sigma}$
\cite{Defenu17}.
Here, the symbol $\eta_{SR}(D)$
denotes the critical exponent 
for the short-range model in $D$ dimensions.
The explicit expressions for $\eta_{SR}(D)$  and $z(\sigma)$ are 
unclear. 
We resort to the approximate relations,
$\eta_{SR}=0$ and $z=\sigma/2$ (\ref{z_formula}),
which are validated in the large-$D$ (small-$\sigma$)
regime \cite{Defenu17}.
As mentioned above, these formulas
lead to
the closed expression, Eq.  (\ref{sDrelation}).

\section{\label{section3}
Summary and discussions}

The
quantum spin-$S=1$ Ising chain with the long-range interactions
(\ref{Hamiltonian})
was investigated numerically.
So far, as for the classical counterpart,
a thorough investigation has been made.
Aiming to elucidate the quantum nature of this problem,
we shed light on its 
low-energy spectrum,
namely,
the single-magnon and
bound-state masses, $m_{1,2}$, respectively, in the ordered phase.
For that purpose,
we employed the exact diagonalization method,
which enables us to calculate $m_{1,2}$ directly.
As a result, we 
obtained the $\sigma$-dependent 
scaled bound-state mass, $m_2/m_1$,
for various values of the algebraically-fall-off exponent, $\sigma$.

Thereby,
based on the 
$\sigma \leftrightarrow D$ relation (\ref{sDrelation}),
we obtained the result $m_2/m_1=1.845(10)$ for $D=3$.
Our result is to be compared with the
preceding results,
$m_2/m_1=1.83(3)$,
$1.81$, 
$1.84(1)$,
and $1.82(2)$
via the Monte Carlo \cite{Caselle99}, series expansion
\cite{Dusuel10}, exact diagonalization \cite{Nishiyama14},
and non-perturbative-renormalization-group 
\cite{Rose16}
methods,
respectively.
Hence,
it is indicated that
down to $D=3$,
the validity of
  super-universality 
\cite{Angelini14} is retained even for
the spectral properties such as $m_2/m_1$.
The magnon-bound-state hierarchy
is relevant to
the glueball spectrum for the gauge field theory
\cite{Agostini97,Fiore03}.
It would be intriguing  that such high-energy 
phenomenology is explored
 \cite{Svetitsky82} 
for the magnetic materials \cite{Coldea10}
 with finely-tunable
\cite{Britton12,Islam13,Richerme14}
long-range interactions.

%

%
%
%
%

\end{document}